\documentclass[a4paper,12pt]{article}
\usepackage{graphicx} 
\usepackage{comment}
\usepackage{setspace} 

\date{\empty}

\title{
\begin{flushleft}
{\large \bf
Determination of hydrogen concentration in solids by transmission ERDA under 
nuclear-elastically enhanced recoiling of H by 8 and 9 MeV He  
}
\end{flushleft}
}
\begin{document}
\maketitle
\vspace{-15mm}
\author{\noindent 
Hiroshi Kudo$^{1}$
\footnote{Corresponding author\, {\it E-mail address}: kudo@tac.tsukuba.ac.jp} 
, Masanori Kurosawa$^{2}$, Hiroshi Naramoto$^{1}$, Masao Sataka$^{1}$, Satoshi Ishii$^{1}$, Kimikazu Sasa$^{1, 3}$ and Shigeo Tomita$^{3}$
}
\\\\
$^{1}${\it Research Facility Center for Science and Technology, University of Tsukuba, 1-1-1 Tennodai, Tsukuba, Ibaraki 305-8577, Japan}\\
$^{2}${\it Faculty of Life and Environmental Sciences, University of Tsukuba, 1-1-1 Tennodai, Tsukuba, Ibaraki 305-8572, Japan}\\
$^{3}${\it Faculty of Pure and Applied Sciences, University of Tsukuba, 1-1-1 Tennodai, Tsukuba, Ibaraki 305-8571, Japan}\\\\
ABSTRACT\\
\setstretch{1.5}

\indent Hydrogen concentrations in thin self-supporting samples of polyphenylene sulfide and 
muscovite have been determined by nuclear-elastic recoil detection analysis of transmission layout. 
The analysis procedure is based only on the database of stopping power and recoil cross section 
for material analysis, without using any reference sample of known H content. 
For the polyphenylene sulfide sample, the determined value of 
$(2.87\pm 0.26)\times 10^{22}\,{\rm H}/$cm$^{3}$ is in good agreement with the calculated 
value of $3.01 \times 10^{22}\,{\rm H}/$cm$^{3}$.  
For the muscovite sample, the H concentration originating each from bound water 
and absorbed water is uniform over the entire thickness of the sample. 
The determined concentration $(9.43 \pm 0.75) \times 10^{21}$ H$/$cm$^{3}$ of the muscovite 
agrees excellently with the value of $9.36 \times 10^{21}$ H$/$cm$^{3}$ obtained from other  
quantitative analyses typically applied for minerals. 
The present results demonstrate the capability of accurate determination of H contents in 
materials and minerals by transmission ERDA. 

\newpage

\section{\large Introduction}\label{intro}

Knowledge about behavior of hydrogen in solids is of scientific and also of practical importance 
since hydrogen can seriously affect physical and, especially, mechanical properties of materials \cite{Con2018}.   
Also, hydrogen in solids attracts geological interest since geological samples typically contain H in the form of 
water\,(H$_{2}$O) or hydroxyls\,(OH) \cite{Swe1997} \cite{Rae2008} \cite{Bur2009}. For example,  it is certainly 
helpful to discuss the origin of the lunar water and the earth's sea via hydrogen analyses of glasses and 
minerals in lunar rocks \cite{Ste2019} \cite{Lin2021}. 

For analysis of hydrogen in solids, possible experimental methods, in particular, of non-destructive type are 
extremely restricted because analysis techniques based on electronic transitions of atoms are 
hardly applicable in this case.   
A notable and unique feature of elastic recoil detection analysis (ERDA) using a fast ion beam is non-destructive 
detection of light atoms in a solid sample \cite{Handbook1995}. For analysis of hydrogen in 
thin self-supporting samples, a useful method is ERDA of transmission layout (T-ERDA), in which recoil H 
resulting from hard collisions with He or heavier projectiles is energy-analyzed in the direction 
close to 0$^{\circ}$ after passing through the sample.
For example, Tirira and coworkers carried out profiling of H in thin self-supporting samples by T-ERDA 
using 2--3 MeV He \cite{Tir1990a} \cite{Tir1990b}. At about the same time, elastic recoil coincidence spectroscopy 
(ERCS), which is a variation of T-ERDA, has been developed \cite{Hof1990}.  Later, ERCS using proton--proton 
scattering was applied to observe the fluid inclusion in minerals \cite{Fur2003}, localized H distribution in 
diamond \cite{Rei2004}, and H concentration in clinopyroxene crystals \cite{Wei2018}. In the studies mentioned above, 
the workers obtained higher spatial resolutions than for the typical case of neutron tomography \cite{Gri2014}, 
which is a competitive analysis technique with ERDA. 

Previously, we observed 3D distribution of hydrogen bubbles in a H-introduced Al foil by T-ERDA  \cite{Yam2019b}.  
Actually, such 3D analysis of hydrogen can be carried out by taking advantage of the enhanced recoil cross section 
as large as $\sim$2\,b/sr due to the nuclear elastic interaction of  8--10\,MeV $^{4}$He with H \cite{iba}.  
This allows sufficient count rate of recoil H needed for the 3D observation by employing  the collimated 
10--100 pA He$^{2+}$ beam of the $\mu$m size.

In our developing stage of nuclear-elastic T-ERDA,  the main factors determining the spatial and depth 
resolutions typically in the $\mu$m range were studied using test samples \cite{Kud2021}. 
Further fundamental studies are needed to establish this method as a standard technique of hydrogen analysis. 
For this reason, we have carried out T-ERDA using samples of PPS 
[polyphenylene sulfide (C$_{6}$H$_{4}$S)$_{\rm n}$] and natural muscovite. The former is a synthetic resin with 
well-known atomic composition, while the latter sample has been highly characterized by other analysis approaches, 
as will be noted in \S\ref{exp}. Rigorous analysis of hydrogen content in these samples is of practical importance 
since these are often used as reference samples for hydrogen analysis by ERDA \cite{Ros2015} \cite{Kin2021}. 
The main interest of the present work is the reproducibility of the known concentrations of hydrogen in 
the samples without relying on any reference sample to determine the absolute H concentrations. 
Therefore, the present analysis method is based fully on the open database of stopping power 
and recoil cross section for material analysis using ion beams. 

\section{ \large Experiments }\label{exp}

In the present experiments, the measured samples are stacked PPS--Al layers and muscovite, 
which are on the Al foil of 50 $\mu$m thickness to prevent the incident He from entering the particle detector. 
The PPS--Al stacked samples are those used in the previous work \cite{Kud2021}, which 
are abbreviated as PPS/Al(50), Al(10)/PPS/Al(50), and Al(20)/PPS/Al(50), where the numbers in 
the parentheses denote the thicknesses in $\mu$m. The thicknesses of the PPS film and the Al foils were 
determined by measuring the weight and area, assuming the densities of 2.70 and 1.35 g/cm$^{3}$ 
for Al and PPS, respectively. The nominal purity of the Al foils is 99.99\%. The measurement reproduced the nominal 
Al thicknesses of 10, 20, and 50 $\mu$m within the error of $\pm$2\%. The measured PPS thickness is 
$1.35 \pm 0.05\, \mu$m, which is thin enough to be used as a localized H marker. Furthermore, the PPS film was 
characterized by Rutherford backscattering spectroscopy (RBS) using 2 MeV He, which showed no discernible 
inhomogeneity of the PPS film, according to the composition analysis of the RBS spectra (S and C) applying 
the commonly used SIMNRA code. 

The muscovite sample was prepared by cleaving the bulk muscovite which was collected from a granitic pegmatite 
rock at Ishikawa, Fukushima Prefecture, Japan. The precise thickness of the muscovite sample on the order of 
10 $\mu$m can be determined afterward from the analysis of the T-ERDA spectrum. 
The density of the muscovite 2.83 g/cm$^{3}$ was measured with a high precision electronic densimeter. 
The chemical composition of the muscovite was determined with scanning electron microscope  
coupled with energy dispersive X-ray spectroscopy (SEM-EDS), while the water content of the muscovite was 
determined by Karl Fischer titration. Averaging the four values of water content 
obtained from four measurements, we obtained 4.53$\pm$0.06 and 0.40$\pm$0.01\,wt\% for 
bound water H$_{2}$O($+$) and absorbed water H$_{2}$O($-$), respectively. 
The analysis results are summarized in table \ref{table-1}, which has well reproduced the published 
data \cite{Uda1973}. It is noted that the critical temperature of 200$^{\circ}$C results from the 
fact that at higher temperatures appreciable weight loss with endothermic reaction occurs \cite{Gug1987}. 
\begin{table}
\caption {Measured chemical composition and thereby obtained atomic composition 
of the muscovite collected from Ishikawa. The content of H$_{2}$O($-$) extracted by heating up 
to 200$^{\circ}$C is considered as absorbed water molecules, 
while H$_{2}$O($+$) extracted by heating from 200 to 1000$^{\circ}$C 
should result from the process of dehydroxylation.  
For this reason, the sum of the chemical composition excluding H$_{2}$O($-$) 
is normalized to 100wt\%. The notations H($-$) and H($+$) are used similarly. }
 \vskip1.5mm
\begin{center}
 \begin{tabular}{lrcrrr}
 \hline\hline 
  & & \\[-8pt]
 Chem. com. & wt\% & \hspace{14mm} Atom. com.  &  at\%  & mass\% \\[1.5mm] 
\hline 
   & & \\[-5pt]
  SiO$_{2}$ & 46.72 & \hspace{14mm} Si & 14.70 & 21.75 \\[1.5mm]
  Al$_{2}$O$_{3}$  & 33.73 & \hspace{14mm} Al & 12.51 & 17.78 \\[1.5mm]
  K$_{2}$O & 10.93 & \hspace{14mm} K & 4.39 & 9.04 \\[1.5mm]
  TiO$_{2}$ & 0.07 & \hspace{14mm} Ti & 0.02& 0.04 \\[1.5mm]
  FeO & 3.07 &\hspace{14mm} Fe & 0.81 & 2.38 \\[1.5mm]
  MnO & 0.09 &\hspace{14mm} Mn & 0.02 & 0.07 \\[1.5mm]
  MgO & 0.21 & \hspace{14mm} Mg & 0.10 & 0.13 \\[1.5mm]
  Na$_{2}$O & 0.65 & \hspace{14mm} Na & 0.40 &  0.48  \\[1.5mm]
  H$_{2}$O ($+$) & 4.53  & \hspace{14mm} H ($+$) &9.51 & 0.50 \\[1.5mm]
  H$_{2}$O ($-$) & 0.40 & \hspace{14mm} H ($-$) &0.84 & 0.04 \\[1.5mm]
   &  & \hspace{14mm} O & 56.70 & 47.79 \\[1.5mm]
  \hline \hline
 \end{tabular}
\end{center}
\label{table-1}
\end{table}

\begin{figure}[h] 
\begin{center}
\includegraphics[width=125mm, bb=94 624 515 773]{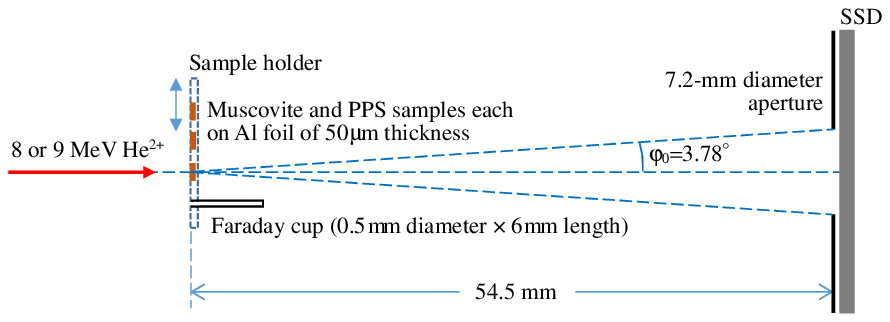}
\end{center}
\caption{Schematic diagram of the experimental setup for T-ERDA. }
\label{fig-1}
\end{figure}

Figure \ref{fig-1} shows the experimental setup similar to the previous case \cite{Kud2021}, except for the 
slightly narrow acceptance angle of the surface-barrier silicon detector (SSD), $\varphi_{0}=3.78^{\circ}$. 
The energy resolution of the detector was 33 keV, which is sufficient for the present T-ERDA.
The experiments were carried out under the pressure of $2 \times 10^{-5}$ Pa. 
The incident He energies of 8 and 9 MeV were chosen for effective use of the extremely enhanced 
He-H differential recoil cross section in the forward direction. The estimated angular spread of 
the incident He$^{2+}$ beam from the UTTAC 6MV tandem accelerator is $\sim$0.2$^{\circ}$, 
which is much less than $\varphi_{0}$. The He$^{2+}$ beam was focused to $\sim$30$\times 30\,\mu$m$^{2}$ 
on the samples. The energy-measuring system for H was calibrated 
with 5.48 MeV $\alpha$ particles emitted from $^{241}$Am. 

The beam current was adjusted within the range of 150--300 pA, depending on the sample, 
to obtain the count rate of 200 counts/s or less for recoil H. This assures negligible counting loss of the H signals 
in the present experiments. The beam current was frequently monitored during the measurements by 
the Faraday cup attached to the sample holder, shown in figure \ref{fig-1}.  The integrated beam current 
(incident charge) for 1 minute was measured repeatedly with a current integrator, from which the uncertainty of the 
incident charge $\pm 5$\% was obtained. The integrated beam current for 1 minute was also measured 
before and after accumulating every hydrogen spectrum for 10 minutes.  
This procedure was repeated two or three times and,  after confirming the reproducibility, the two or three hydrogen spectra  
were summed up to obtain the final data. 

Under fixed beam conditions, we have repeated to measure H spectra several times at several fixed beam positions 
on the samples and confirmed that there is only negligible difference among 
these spectra. This indicates not only negligible loss of H from the samples by the He irradiation, 
but also uniformity of atomic composition and thickness of the samples.
 
\section{ \large Results and discussion }\label{discuss}
According to the previous study \cite{Kud2021}, the present T-ERDA can be discussed 
using the straight-path approximation for the trajectories of He and recoil H. 
Also, the enhanced energy loss for the acceptance angle of $\varphi_{0}=3.78^{\circ}$ relative to $0^{\circ}$ is 
negligibly small since the path length is enhanced only by $1/\cos{\varphi_{0}}-1$, i.e.,  0.22\%. 
It is notable that the angular spread of recoil H after passing through the present sample  causes 
negligible change in the H yield. This is because the differential recoil cross section changes slowly 
with recoil angle in the narrow forward direction under consideration, hence the H yield is unchanged 
as a whole even when the recoil directions of recoil H are smeared and mixed with each other by 
the multiple scattering suffered in the sample. In a representative case, for example, the angular spread 
of 5.12 MeV H (recoil H by 8 MeV He) after passing through an Al foil of 50 $\mu$m is $\pm 1.2^{\circ}$ 
\cite{Kud2021}, which corresponds to only $\pm 1$\% change in the value of differential recoil cross section 
at the recoil angle of 3.78$^{\circ}$ \cite{iba}. It is of essential importance to note that the analysis of 
observed energy spectra of recoil H requires to know, in advance, the stopping powers for incident 
He and recoil H, which are evaluated from the density and the atomic composition of the sample. 
This situation might seem logically inconsistent in the determination of H concentration. 
In most cases, however, a possible H content lies in a relatively limited range.  Hence, hydrogen in the sample 
makes only minor contribution to the stopping powers evaluated for a fixed value of the known target density, 
and therefore, we can determine yet unknown H concentration in the sample. 
The following analysis of the experimental data is based on the physical backgrounds mentioned above. 

\subsection{\normalsize PPS}
\label{sec-PPS}

\begin{figure}[h] 
\begin{center}
\includegraphics[width=120mm, bb=99 367 527 670]{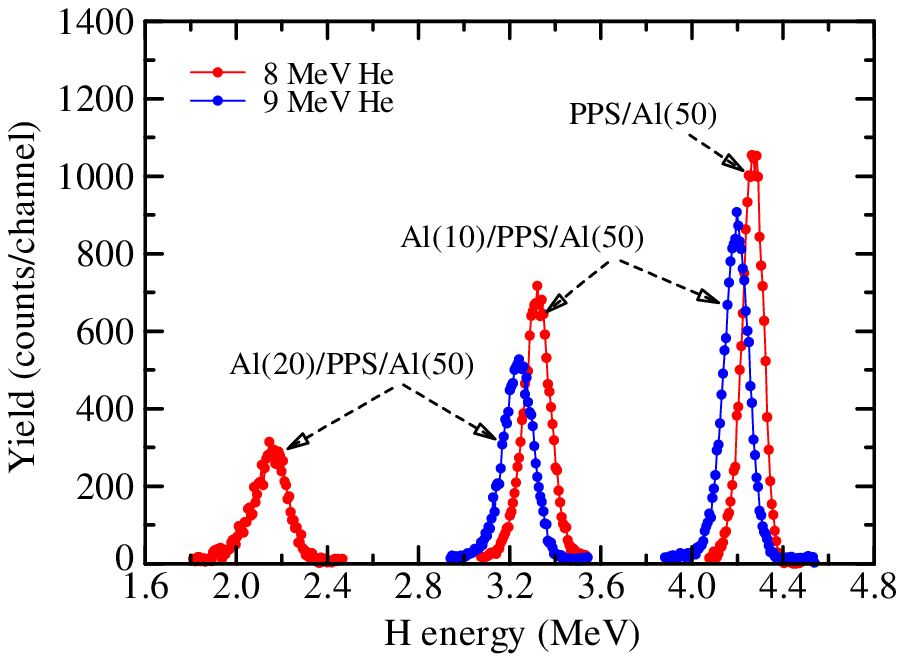}
\end{center}
\caption{T-ERDA spectra for incidence of 8 and 9 MeV He on the three PPS samples, shown 
for the same number of incident He$^{2+}$, $2\times 10^{11}$. Note that H energy of 6.683 keV corresponds to 
1\,channel width on the horizontal axis.}
\label{fig-2}
\end{figure}

Figure \ref{fig-2} shows the energy spectra of recoil H measured using 8 and 9 MeV 
He$^{2+}$ for the stacked PPS--Al samples. The absence of PPS/Al(50) data for 9 MeV He is because  
the incident He can pass through the sample and certainly damage the detector. 
The sharp peaks in the spectra originate from recoil H in the PPS film. The peak energies for 
8 MeV He$^{2+}$ agree with the previously observed values \cite{Kud2021}. These peak energies can be 
reproduced with a discrepancy of only 0.6\% by the evaluations of energy losses in PPS and Al, 
in which the straight path approximation and the stopping powers by SRIM are employed. 
On the other hand, the similar evaluations for 9 MeV underestimate the peak energies by 1.2\%, 
corresponding to $\sim$40 keV, which is not negligible in comparison with the energy resolution 
of the detector 33 keV, as noted earlier. Nevertheless, such discrepancies with the peak energies do not 
affect the present analysis of the experimental data. 

Since the PPS thickness (1.35\,$\mu$m) cannot be resolved with the 
depth resolution, 1.5--1.6 $\mu$m, of the present T-ERDA \cite{Kud2021}, 
the areas under the peaks are considered to determine the H concentration in PPS. 
In this case, the slight and approximately uniform background yield originating probably from He--Al 
nuclear reactions must be subtracted from every spectrum.
For each spectrum, the background was subtracted by assuming the straight-line background connecting 
the ends of low- and high-energy tails of the H peak, which are well recognized in the spectrum. 
In every case, the area under the H peak includes the background yield of 2--10\%. 
Since the background yield determined in such manner has typical uncertainty of $\pm 10$\%, 
the uncertainty of the H yield after background subtraction should be at most 10\%$\times$10\%, i.e., 1\%. 
The data for the five different experimental conditions, shown in table 2, each providing the H concentration 
in PPS, are useful to confirm the consistency and reliability of the present analysis procedure. 

\begin{figure}[h] 
\begin{center}
\includegraphics[width=120mm, bb=77 385 491 636]{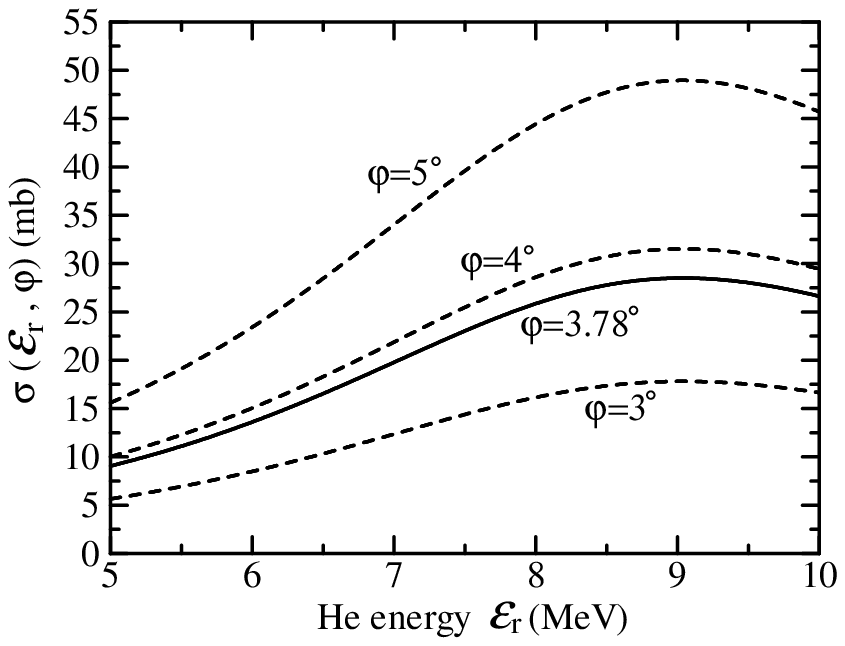}
\end{center}
\caption{ $\sigma ({\cal E}_{\rm r}, \varphi)$ calculated using the numerical table of differential recoil cross 
sections given by IBANDL \cite{iba}. }
\label{fig-3}
\end{figure}

The integrated H yield under the peak $Y_{\rm peak}$ is related to the H concentration $n$ by
\begin{eqnarray}
Y_{\rm peak}=N n t \times \int \frac{d \sigma}{d \Omega}\, d\Omega \;, 
\label{eq-11}
\end{eqnarray}
where $N$ is the number of incident He, $t = 1.35\,\mu$m is the thickness of the PPS sample, 
$d \sigma / d \Omega$ is the He--H  differential recoil cross section 
in the laboratory frame, which depends on the He energy at recoil ${\cal E}_{\rm r}$ and the solid angle 
$\Omega$, and the integral is performed for the solid angle of acceptance of the detector. 
${\cal E}_{\rm r}$ can be calculated using the stopping powers for He. 
Equation (\ref{eq-11}) leads to 
\begin{eqnarray}
n=\frac{Y_{\rm peak}}{Nt\, \sigma ({\cal E}_{\rm r}, \varphi_{0})} \;,
\label{eq-22}
\end{eqnarray}
where the cross section $\sigma ({\cal E}_{\rm r}, \varphi_{0})$ is expressed as 
\begin{eqnarray}
\sigma ({\cal E}_{\rm r}, \varphi_{0}) =  \int \frac{d \sigma}{d \Omega}\, d\Omega 
=\int_{0}^{\varphi_{0}}\frac{d \sigma}{d \Omega}\, 2 \pi\sin{\varphi}\, d\varphi \;. 
\label{eq-33}
\end{eqnarray}
where $\varphi$ is the polar angle with respect to the He beam axis, and $\varphi_{0}=3.78^{\circ}$, as noted in \S\ref{exp}. 
Figure \ref{fig-3} shows $\sigma ({\cal E}_{\rm r}, \varphi)$ for the parameter ranges of interest in the present analysis, 
which was calculated using the numerical table of differential recoil cross sections given by IBANDL \cite{iba}. 
At ${\cal E}_{\rm r}=9$ MeV, $\sigma ({\cal E}_{\rm r}, \varphi)$ is maximal, implying that  
the highest efficiency for collecting H signals is obtained for the present setup of $\varphi=\varphi_{0}\,(=3.78^{\circ})$. 
Nevertheless, the background yield increases with increasing the He energy and therefore 
the He energy must be chosen by considering the balance between the efficiency and sensitivity  
required for hydrogen analysis of a given sample. 
Comparing the values of $\sigma ({\cal E}_{\rm r}, \varphi)$ for $\phi=3.78^{\circ}$ and 4$^{\circ}$, we notice the 
appreciable difference by $\sim$10\% only for the small angular change of 0.22$^{\circ}$. This is the reason 
that special care was taken for the acceptance angle of the detector in the experimental setup.

Table \ref{table-2} shows the determined values of the H concentration of PPS, for which the experimental 
data used are in the ranges of $\Delta Y =$(1.4--2.4)$\times 10^{4}$ and $N =$(1.2--2.8)$\times 10^{11}$. 
For the case of 9 MeV He on the Al(10)/PPS/Al(50) sample, for example, the concentration of $2.68 \times 10^{22}$ 
H/cm$^{3}$ results from $\Delta Y =$2.311$\times 10^{4}$ after 
subtracting the background yield of $1.55 \times 10^{3}$, $N =2.588 \times 10^{11}$, and 
${\cal E}_{\rm r}=7.77$ MeV which corresponds to $\sigma = 24.65$ mb (see figure \ref{fig-3}).  
Note that the energy loss of He and recoil H in the PPS sample is negligibly small in the determination of the 
H concentration, as is obvious from the small energy differences in the values of ${\cal E}_{\rm r}$, shown in table \ref{table-2}. 
\begin{table}
\caption {H concentration in PPS determined under different measurement 
conditions. The He energy at recoil ${\cal E}_{\rm r}$ is shown in terms of the central energy 
with $\pm$difference resulting from the finite thickness of PPS. 
The estimated uncertainty in the values of H concentration is $\pm$8\%.  
The rightmost column shows the ratio of the determined concentration to the calculated concentration 
(3.01$\times$$10^{22} {\rm H}/$cm$^{3}$) from the atomic composition of PPS.  }
 \vskip1.5mm
\begin{center}
 \begin{tabular}{cccccr}
 \hline\hline 
  & & \\[-5pt]
  Sample & He energy & He energy  &  H concentration &  Ratio to \\
   &  ${\cal E}_{\rm in}$(MeV) & ${\cal E}_{\rm r}$(MeV) &  ($10^{22} {\rm H}/$cm$^{3}$) 
   & 3.01$\times$$10^{22} {\rm H}/$cm$^{3}$  & \\[1.5mm] \hline 
   & & \\[-5pt]
  PPS$/$Al(50) & 8 & 7.95$\pm$0.05 &2.64 & 0.88 \\[1.5mm]
  Al(10)$/$PPS$/$Al(50) & 8 & 6.66$\pm$0.06 & 3.17 & 1.05 \\[1.5mm]
  Al(10)$/$PPS$/$Al(50) & 9 & 7.77$\pm$0.05 & 2.68 & 0.89 \\[1.5mm]
  Al(20)$/$PPS$/$Al(50) & 8 & 5.21$\pm$0.07 & 2.94 & 0.98 \\[1.5mm]
  Al(20)$/$PPS$/$Al(50) & 9 & 6.44$\pm$0.06 & 2.90 & 0.96 \\[2.5mm]
  \hline \hline
 \end{tabular}
\end{center}
\label{table-2}
\end{table}

The uncertainty of the determined values of H concentration stems mainly from five sources: (i) the $\pm 5$\% error 
in $N$ associated with the integrated current, as is mentioned in \S\ref{exp}, (ii) a possible error of $\pm 4$\% 
in $\sigma ({\cal E}_{\rm r}, \varphi)$ resulting from the $\pm 0.07^{\circ}$ uncertainty of the acceptance angle 
of the detector, (iii) a possible error of $\pm 4$\% associated with the measured value of $t$, (iv) 
the statistical error of $\Delta Y \simeq 1$\%, and (v) a 1\% error due to  
the background subtraction, discussed earlier. 
Because of no interaction among these five sources, the total uncertainty is estimated as  
$\pm (5^{2}+4^{2}+4^{2}+1^{2}+1^{2} )^{1/2} \simeq \pm 8$\% by assuming 
the convolution characteristic of Gaussian distributions.  
It should be noted that in the literatures there is a discrepancy of $\sim$$\pm 5$\% between 
the experimental and theoretical recoil cross sections for the energy and angular ranges of interest 
\cite{Pus2004}\cite{Hup2014}. Even if this discrepancy is taken into account, only minor change 
arises in the above estimate, i.e., from 8 to 9\%. 

The H concentration in PPS calculated from the nominal density of 1.35 g/cm$^{3}$
and the atomic composition of (C$_{6}$H$_{4}$S)$_{\rm n}$, using Avogadro's constant,
 is $3.01 \times 10^{22}\,{\rm H}/$cm$^{3}$. 
The ratios of the determined value to the calculated value, shown in table \ref{table-2}, 
are approximately equal to 1.  Averaging the five determined values, 
we may conclude $(2.87\pm 0.26)\times 10^{22}\,{\rm H}/$cm$^{3}$ for the H concentration of PPS. 
This value agrees well and consistently with the calculated one, which demonstrates the 
validity of the present analysis. 

\subsection{\normalsize Muscovite}

\begin{figure}[h] 
\begin{center}
\includegraphics[width=120mm, bb=74 387 495 698]{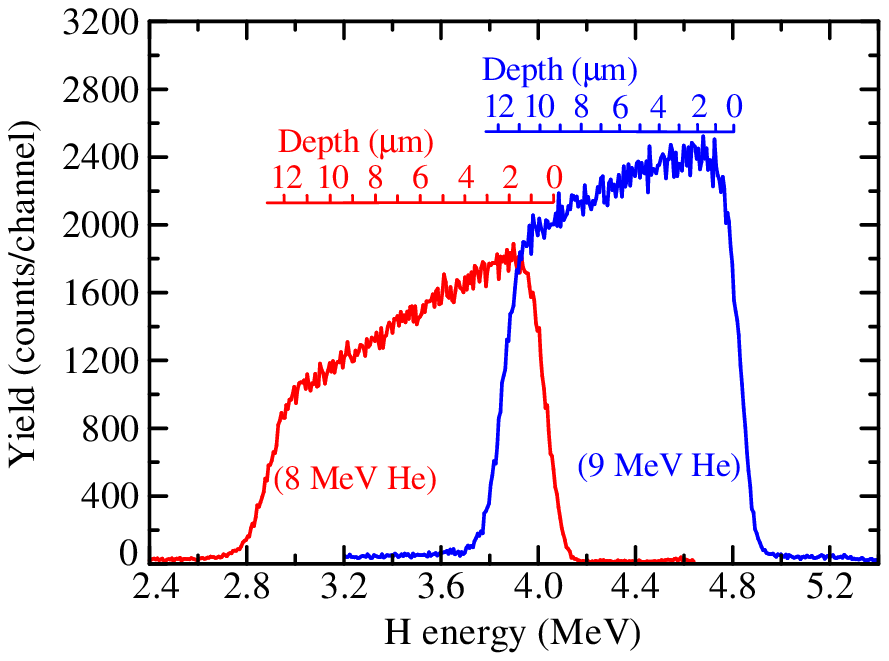}
\end{center}
\caption{T-ERDA spectra of recoil H from the muscovite sample, shown for the same number 
$1.05 \times 10^{12}$ of incident  8 and 9 MeV  He$^{2+}$. 
Note that H energy of 6.617 keV corresponds to 1\,channel width on the horizontal axis.} 
\label{fig-4}
\end{figure}

Figure \ref{fig-4} shows T-ERDA spectra of recoil H from the muscovite sample, which were 
measured with 8 and 9 MeV He$^{2+}$. The counts of the spectrum yield for 9 MeV have been 
multiplied by 1.103 to compare the two spectra for the equal number of incident He$^{2+}$, 
$1.05 \times 10^{12}$ for the 8 MeV case. The narrower and higher spectrum of recoil H for 9 MeV 
results from the reduced energy loss of He and H in the sample and from the increased 
recoil cross sections, compared with the 8 MeV case.  

For evaluation of the stopping powers of the muscovite, its density 2.83 g/cm$^{3}$ and 
the atomic composition shown in table \ref{table-1} are the input parameters of the SRIM code \cite{Zie2013}. 
Compared with pure muscovite of 2.9 g/cm$^{3}$, which consists of K, Al, Si, O, and H 
[\,KAl$_{2}$\,AlSi$_{3}$\,O$_{10}$\,(OH)$_{2}$\,], there are 
several substituted atoms in the muscovite sample. However, such difference, including the different 
H contents, hardly affect the stopping powers. Actually, the stopping powers of the muscovite sample 
for He and H in the energy ranges under consideration are less than those of 
pure muscovite by only 0.2\%. 

In the straight-path approximation, the measured H energy $E$ can be converted to the depth $z$ 
where the recoil occurs in the muscovite of thickness $L_{\rm mus}$. We start with writing down the He energy 
at recoil, which was introduced in \S\ref{sec-PPS}, i.e.,   
\begin{eqnarray}
{\cal E}_{\rm r}= {\cal E_{\rm in}}-\int_{0}^{z} S_{\rm mus-He}\, dz \:, 
\label{eq-1}
\end{eqnarray}
where ${\cal E_{\rm in}}=8$ or 9\,MeV is the kinetic energy of incident He in the present case, and the notation 
$S_{\rm mus-He}$ indicates the stopping power of muscovite for He.
Since recoil H passes through the Al foil of thickness 
$L_{\rm al}=50\, \mu$m before entering the detector, the relation between $E$ and  $z$ is written in terms of 
the integral along $z$ which is also the beam axis of incident He, i.e.,  
\begin{eqnarray}
E=\alpha \, {\cal E}_{\rm r} - \int_{z}^{L_{\rm mus}} S_{\rm mus-H}\, dz 
- \int_{L_{\rm mus}}^{L_{\rm mus}+L_{\rm al}} S_{\rm al-H}\, dz\:,
\label{eq-2}
\end{eqnarray}
where $\alpha=4 M_{\rm H}  M_{\rm He}/(M_{\rm H}+ M_{\rm He})^{2}=0.64$ with $M_{\rm H}$ and $M_{\rm He}$ 
being the atomic masses of H and He, respectively, and the stopping powers are written with the same notation 
as in equation\,(\ref{eq-1}). 
The stopping powers in equations\,(\ref{eq-1}) and (\ref{eq-2}) depend on the kinetic energy of He or H, 
and therefore, on $z$, hence they increase as $z$ increases (with losing the kinetic energies) in the 
energy ranges of He and H under consideration.  
The first term on the right side of equation\,(\ref{eq-2}) represents the energy of recoil H induced at $z$, 
while the second and the third represent the energy loss of H in the outgoing path. 

\begin{figure}[h] 
\begin{center}
\includegraphics[width=120mm, bb=83 353 499 697]{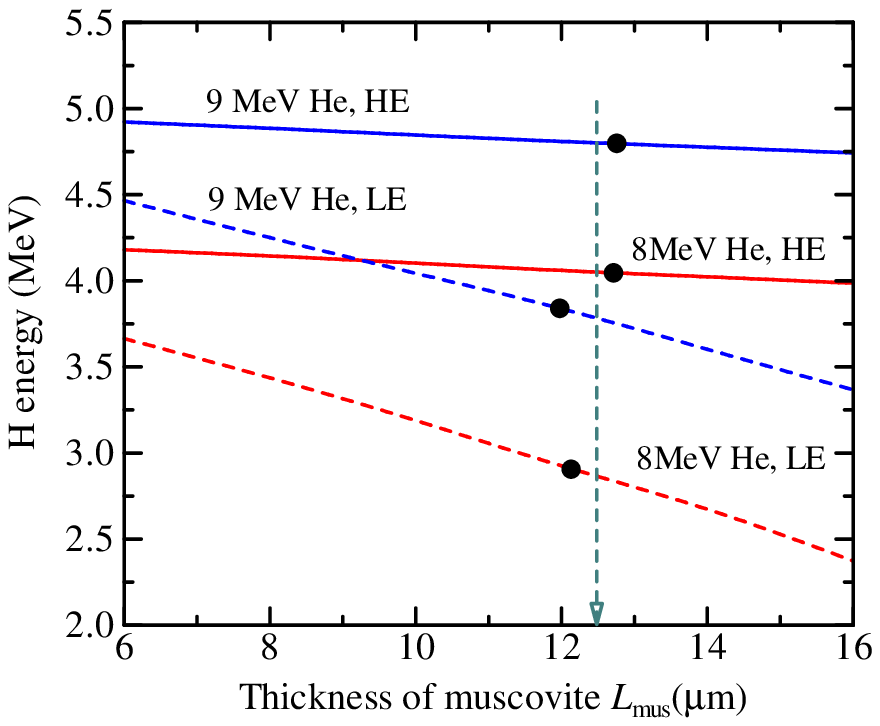}
\end{center}
\caption{Calculated high- and low-energy edges (denoted as HE and LE, respectively) of the H-energy 
spectrum as a function of the thickness of muscovite. The black dots on the curves 
indicate the  HE and LE values of the measured spectra shown in figure \ref{fig-4}, from which the most 
probable thickness of $12.5\, \mu$m (the green dashed arrow) has been determined. }
\label{fig-5}
\end{figure}

For a given value of $L_{\rm mus}$,  $E$ decreases  with increasing $z$ from $z=0$  
because of the larger stopping power for He than for H. 
It is obvious that $E$ at $z=0$ decreases with increasing $L_{\rm mus}$ and, accordingly, $L_{\rm mus}$ can be 
determined from such a condition that the calculated value of $E$ at $z=0$ reproduces the high-energy edge of 
the measured spectrum. Similarly, the low-energy edge of the measured spectrum can also be used to 
determine $L_{\rm mus}$, which allows to reconfirm the value of $L_{\rm mus}$ obtained from the high-energy edge. 
Actually, the high- and low-energy edges are given by the H energies at half-heights of the right and left steps of 
the spectra shown in figure \ref{fig-4}. This procedure is illustrated in figure \ref{fig-5}, thereby we have determined 
$L_{\rm mus} =12.5 \pm 0.5\,\mu$m using the values of stopping powers by SRIM.  
The depth scales in figure \ref{fig-2} have been obtained accordingly. These depth scales and the smeared
high- and low-energy edges of the spectra indicate that the depth resolution of the present T-ERDA is $\sim$2 $\mu$m. 

The depth-dependent H concentration in the muscovite sample $n(z)$ can be obtained from the spectra in 
figure \ref{fig-4}. When the $z$ axis is divided by a short interval of $\Delta z$, the H yield contributed from 
the thickness range of  $z-\Delta z/2$ to $z+\Delta z/2$ is expressed as 
\begin{eqnarray}
\Delta Y(z) = N\,  n(z)\,\Delta z \int \frac{d \sigma}{d \Omega}\, d\Omega \;, 
\label{eq-3}
\end{eqnarray} 
where $N$ is the number of incident He, $d \sigma / d \Omega$ is the He--H  differential recoil cross section 
in the laboratory frame, which depends on the He energy at recoil ${\cal E}_{\rm r}$ and the solid angle 
$\Omega$, and the integral is performed for the solid angle of acceptance of the detector.  
Notably, equation\,(\ref{eq-3}) is on the basis of the fact that the counting efficiency of the detector for recoil H is 
effectively 100\%, which has been assured, although indirectly, from the analysis results of PPS in \S\ref{sec-PPS}. 
From equation (\ref{eq-3}), we obtain 
\begin{eqnarray}
n(z)=\frac{\Delta Y(z)}{N\, \sigma ({\cal E}_{\rm r}, \varphi_{0})\,\Delta z} \;, 
\label{eq-4}
\end{eqnarray} 
where the recoil cross section $\sigma ({\cal E}_{\rm r}, \varphi)$ is expressed as 
\begin{eqnarray}
\sigma ({\cal E}_{\rm r}, \varphi) =  \int \frac{d \sigma}{d \Omega}\, d\Omega 
=\int_{0}^{\varphi}\frac{d \sigma}{d \Omega}\, 2 \pi\sin{\varphi}\, d\varphi \;. 
\label{eq-5}
\end{eqnarray}

\begin{figure}[h] 
\begin{center}
\includegraphics[width=120mm, bb=90 441 507 734]{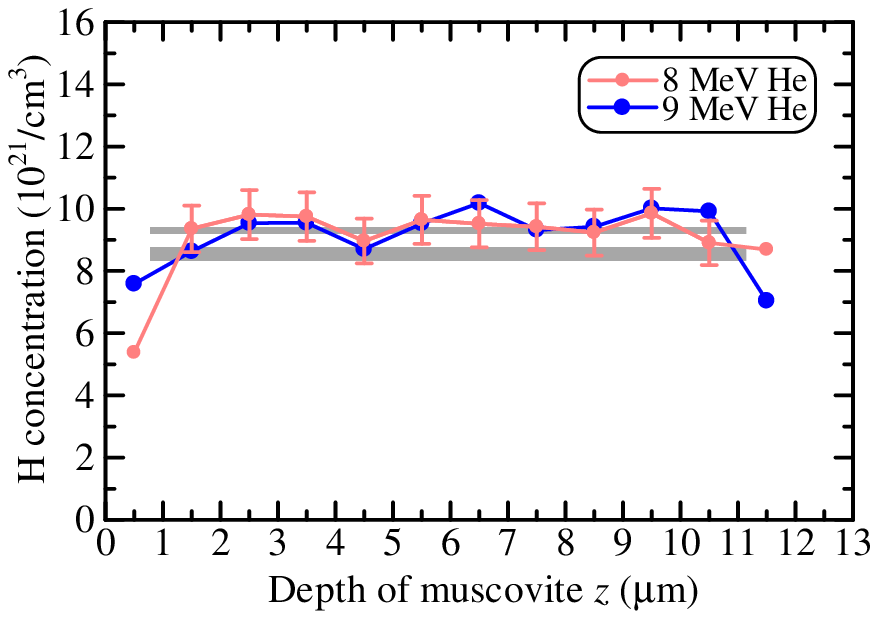}
\end{center}
\caption{H concentration in the muscovite sample, determined by T-ERDA using 8 and 9 MeV He$^{2+}$. 
The uncertainty of $\pm$8\% in the determined values is shown only for the 8 MeV case to avoid a confusing figure. 
The horizontal gray thick lines indicate the calculated concentrations of $9.36\times 10^{21}$ H$/$cm$^{3}$ (upper) 
and $8.61 \times 10^{21}$ H$/$cm$^{3}$ (lower), with and without the absorbed water  
in the muscovite, respectively. Note that their line widths represent the uncertainties of 
the calculated concentrations, see text.} 
\label{fig-6}
\end{figure}

It is straightforward that $n(z)$ for $\Delta z = 1\,\mu$m can be determined using equation (\ref{eq-4}) and  
$\sigma ({\cal E}_{\rm r}, \varphi)$ shown in figure \ref{fig-3}. In this case, $\Delta Y$ for every 
1 $\mu$m depth is obtained from the spectra in figure \ref{fig-4} after subtracting the slightly recognized background.  
Figure \ref{fig-6} shows $n(z)$ determined in this manner. For example, the concentration of 
$n(z)=9.52\times 10^{21}$ H$/$cm$^{3}$ at  $z=6.5\,\mu$m for 8 MeV results from $\Delta Y=20,179$ counts 
for $6 \le z \le 7\,\mu$m, $N=1.054 \times 10^{12}$, and  $\sigma ({\cal E}_{\rm r}, \varphi_{0})=20.10$ mb.

We see in figure \ref{fig-6} that the results of $n(z)$ for 8 and 9 MeV He agree consistently with each other  
at the depth from 1.5 to 11 $\mu$m. Outside this depth range, the slopes near the high- and low-energy 
ends of the spectra should lead to unreliable results. We see the uniformity of $n(z)$ at 
$1.5 \le z \le 11\,\mu$m, hence the average value ${\bar n} = (9.43 \pm 0.75) \times 10^{21}$ H$/$cm$^{3}$ 
is obtained. The uncertainty of ${\bar n}$ has been estimated as $\pm 8$\% in the similar manner to the case 
of PPS, except that the possible error of the PPS thickness is replaced by that of the muscovite thickness, 
$0.5/12.5=4$\%, as noted earlier. 
In figure \ref{fig-6}, the value of ${\bar n}$ agrees well with the H concentration of 
$9.36\times 10^{21}$ H$/$cm$^{3}$ obtained from the data in table \ref{table-1}, using  
the density of the muscovite 2.83 g/cm$^{3}$ and Avogadro's constant. Furthermore. the value of ${\bar n}$ 
cannot be accounted for with the concentration of $8.61\times 10^{21}$ H$/$cm$^{3}$ which is obtained for the 
bound water only. Therefore, we may conclude that the absorbed water is distributed uniformly 
over the entire thickness of the muscovite sample with the present depth resolution of $\sim$2 $\mu$m. 
This is unique information obtained by the nondestructive T-ERDA. 
These analysis results demonstrate the capability of accurate determination of hydrogen contents 
in materials and minerals by the present T-ERDA system. 

\section{ \large Summary and Conclusion }\label{conclude}

Hydrogen concentrations in PPS and muscovite have been determined by T-ERDA without using 
reference samples. For PPS, the determined H concentration of   
$(2.87\pm 0.26)\times 10^{22}\,{\rm H}/$cm$^{3}$ is in good agreement with the calculated concentration of 
$3.01 \times 10^{22}\,{\rm H}/$cm$^{3}$.  For muscovite, the H concentration is uniform over the entire 
thickness of the sample with the depth resolution of $\sim$2 $\mu$m, and the average value 
$(9.43 \pm 0.75) \times 10^{21}$ H$/$cm$^{3}$ 
agrees excellently with $9.36 \times 10^{21}$ H$/$cm$^{3}$ which is obtained from the measured 
chemical composition of the muscovite sample, listed in table \ref{table-1}. 
This also implies that the H concentration originating from absorbed water is uniform in the muscovite, 
like that from bound water constituting the crystal structure of muscovite.  
These analysis results assure reliability of 
T-ERDA under the condition of nuclear elastically enhanced recoiling. 

In the present T-ERDA, the sensitivity to H content is restricted by the background yield, noted 
in \S\ref{sec-PPS}. Obviously, the background yield can be suppressed by replacing the present 
stopping foil (50 $\mu$m thick Al) by a foil of heavier element with higher Coulomb barrier. 
Such a refinement in the experimental setup might be of technical 
importance when the T-ERDA system is used for analysis of  not only small amount of hydrogen, 
but also trace amount of deuterium  introduced in a foil sample. 


\newpage
\hspace{-7.5mm} {\bf ACKNOWLEDGEMENTS}\\
\indent The authors thank the technical staff of UTTAC for operating the 
6MV tandem accelerator for supporting the experiments. 
We are grateful to Toray KP Films Inc. for providing the PPS films, and to 
Toyo Aluminium K.\,K. for providing a set of Al foils of different thicknesses. 

\newpage
\newpage
\newpage
\newpage
\newpage
\newpage

\end{document}